\documentclass[12pt]{article}
\usepackage{graphicx}
\begin{document}

\centerline{\large On the free rotation of a molecule embedded in
helium-4 clusters}

\vspace{2cm}

\centerline{\bf Y. S. Jain$^1$ and S. Dey$^{1,2}$}

\bigskip
\noindent
{\bf $^1$Department of Physics, North-Eastern Hill University, Shillong - 793022, India}

\smallskip
\noindent
{\bf $^2$Department of Basic Sciences, Assam Don-Bosco University, Guwahati - 781017, India}

\vspace{2.5cm}
\begin{abstract}
\bigskip
The fact, that $^4$He atoms on different concentric 
circular paths around the axis of a quantum vortex 
move with identically equal angular momentum, which
represents an important aspect of superfluidity of 
He-II, has been used to discover a model which can 
explain the {\it typical nature} of experimentally 
observed $N$ (number of $^4He$ atoms) dependence of 
the rotational constant ($B$) of the rotor part of 
a cluster M:He$_N$.  It reveals how exactly 
superfluidity is related to the said dependence of 
$B$ on $N$.  We believe that this model, when used with 
simulation techniques, would render results that would 
agree closely with experiments.

\end{abstract}

\vspace{2.5cm}
 

\vspace{2.5cm}
\noindent
\copyright \, \, by authors.
\newpage

\newpage
Ever since the spectroscopic study of SF$_6$ molecule embedded 
in superfluid helium-4 (He-II) was performed by Goyal 
{\it et.al.} \cite{goyal}, 
rotational and vibrational dynamics of several molecules (say, 
OCS, CO$_2$, CO, N$_2$O, HCCCN, {\it etc.}, represented by M) 
embedded in bulk He-II and its nano-droplets and clusters of the 
form M:He$_N$ (where $N = 1, 2, 3, ...$ is the number of He atoms) 
have been extensively investigated.  While, in a break through 
work, Grebenov {\it et.al.} \cite{grebe} found that OCS molecule 
embedded in $^4$He droplets, isolated in the normal phase of 
liquid $^3$He, rotates almost like a free rotor if the droplet has 
about 60 or more $^4$He atoms, systematic experimental study of 
M:He$_N$ clusters have demonstrated non-trivial dependence of their 
rotational constant $B$ (or moment of inertia, $I$) and vibrational 
frequency shift $\Delta\nu$ (of select modes of vibrations) with 
$N$ which concludes that superfluidity of $^4$He atoms has observable 
impact on $B$ of such small clusters as well 
\cite{mcke2, mcke3, mcke4, surin, topic}.  A number 
of theoretical models, such as, (i) super-molecule model \cite{hartm}, 
(ii) two fluid model \cite{grebe, kwon}, (iii) quantum hydrodynamic 
model \cite{calth}, {\it etc.}, have been used to explain the initial 
observations.  Simulation techniques have also been used, recently, 
to explain the phenomenon but with limited success at quantitative 
scale \cite{moron, paesa, paoli, 
paes2, blino, zilli, miur2, miura}.  It is not surprising 
because all these models associate the phenomenon with the 
superfluidity of $^4$He atoms which by itself is not clearly 
understood \cite{ander, tom}.  

\bigskip
The conventional microscopic theory (CMT) \cite{ander, tom} of a 
{\it system of interacting bosons} (SIB), such as liquid $^4$He, 
uses single particle basis (SPB) for its description.  It  
considers that particles occupy different quantum states 
of a single particle placed in a box of volume V of the system 
and these states are described by plane waves ($u_k(r)$ = 
$A\exp{(i{\bf k}.{\bf r})}$ where symbols have their usual meaning); 
in other words a single particle represents the basic unit with an 
assumption that its momentum remains a good quantum number even in 
the superfluid state of the system.  The theory 
concludes that: (1) the state of liquid $^4$He at a temperature ($T$) 
can be identified by the momentum distribution $N(p)$ of its 
particles where different number of particles $N_p(T)$ have 
different $p$, and (2) with the onset of superfluid transition, 
$N(p)$ does not change significantly except for the existence of 
a fraction of particles, $n_{p=0}(T) = N_{p=0}/N$ having $p=0$ 
in LT phase.  Accordingly, even the ground state (G-state) of liquid 
$^4$He has different number of atoms in the states of different momenta, 
{\it viz.}, $N_{p=0}$ in $p$(=$\hbar k) = 0$ state and $N_{p\not=0}$ 
in several states of non-zero momenta, {\bf k}$_1$, {\bf k}$_2$, 
{\bf k}$_3$, ...  ... {\it etc.} (expressed in wave number).  Based 
on different estimates by a large number of theoretical and 
experimental studies, $N_{p=0}$ ($p=0$ condensate) is believed 
to fall around 10$\%$ \cite{ander, enss} leaving $N_{p\not=0}$ 
(non-condensate) to 
about 90$\%$.  In what follows CMT identifies that $p=0$ condensate 
as the origin of superfluidity and related properties of He-II 
and the same view point is attributed to the superfluidity exhibited 
by microscopic systems of $^4$He atoms ({\it viz.}, droplets and 
clustes) too.  The advances in CMT made over the last several years 
\cite{tom} suggest that superfluid phase of liquid $^4$He also has 
(in addition to $p=0$ condensate) pair condensate (similar to Copper 
pair condensation in superconductors) or a composite condensate 
($p=0$ condensate, pair condensate, 3 particle condensate, {\it etc.}) 
as the origin of superfluidity.  

\bigskip
We note that SPB used in CMT not only complicates the process of 
finding different aspects of a SIB, {\it e.g.}, the expectation 
value of inter-particle interaction which becomes infinitely 
repulsive for short distances, but also ignores the reality that 
the states of wave superposition assumed by the particles at 
low $T$ can not be described by $u_k(r)$.  It is well known that: 
(i) two particles (say P1 and P2) in the state of their wave 
superposition are described by $\Psi{(1,2)} = [u_k(r)u_k(r) \pm 
u_k(r)u_k(r)$)] which basically represents a pair of particles 
moving with equal and opposite momenta ({\bf q}, -{\bf q}) 
with respect to their CM which moves with momentum {\bf K} in 
the laboratory frame and (ii) positions ({\bf r}$_1$, and {\bf r}$_2$), 
momenta ({\bf k}$_1$ and {\bf k}$_2$) and energies (${\epsilon}_1$ 
and ${\epsilon}_2$) of two particles as separate entities lose 
their meaning in this state. 

\bigskip
While superfluidity undoubtedly has a relation with the experimental 
observation of the rotation of a molecule embedded in a $^4He$ 
droplet and the non-trivial $N-$dependence of $B$ of M:$^4$He$_N$ 
cluster, it certainly has no relation with $p=0$ condensate
because, as estblished unequivocally by one of us (Jain \cite{jain2}), 
$p=0$ condensate does not exist in the superfluid phase of a SIB.  
Starting with an assumption that the G-state of liquid 
$^4$He has non-zero values of both, $N_{p=0}$ and $N_{p\not=0}$, as 
concluded by CMT, Jain \cite{jain2} finds that: (i) such an $N(p)$ does not 
represent a state of minimum possible energy as expected for the G-state
of every physical system, (ii) all particles in the true G-state of 
a SIB have identically equal enegy ($\varepsilon_o = h^2/8md^2$ with 
$h$ being the Planck constant, $m$ the mass of a particle and $d = 
({\rm V}/N)^{1/3}$) and corresponding non-zero momentum 
($q_o = \pi/d$), and (iii) the real form BEC that exists in a SIB is 
the macroscopic condensation of bosons as ({\bf q}, -{\bf q}) pairs 
in their G-state characterized by $q=q_o=\pi/d$ and $K=0$.   

\bigskip
Motivated by all such observations, one of us \cite{jain1} used 
more realistic pair of particle basis {PPB} to conclude his 
non-conventional microscopic theory (NCMT) which emphasizes a pair 
as the basic unit of the system. The theory not only explains different 
properties of He-II at quantitative scale \cite{jain1, simanta, jain4} 
but also reveals that: (i) particles (G-state) of a SIB have identically 
equal energy ($\varepsilon_o$) and corresponding non-zero $q=q_o=\pi/d$ which 
agrees with a recent study by Jain \cite{jain1}, (ii) they constitute 
a kind of close packed arrangement of their representative wave 
packets (CPA-WP) of identically equal size $\lambda/2 = h/2p = d$, 
(iii) they are allowed to move only coherently in order of their 
locations, obviously, with no relative motion and mutual collision, 
and (iv) they occupy phase ($\phi$) positions which differ by 
$\Delta\phi = 2n\pi$ (with $n=$ 1,2,3, ...).  In addition, the theory 
finds that all the three characteristics of the G-state are retained 
by the superfluid phase over the entire range of temperature from 
$T = 0$ to $T_{\lambda}$ and the entire system assumes a kind of 
collective binding for which it behaves like a macroscopic molecule.

\bigskip
Since $^4$He atoms in M:He$_N$ clusters are confined to a space of 
few \AA \, size, it is evident that each $^4$He atom has non-zero 
energy and corresponding non-zero momentum for the confinement.  This 
undoubtedly proves the absence of $p=0$ condensate in these systems 
and we use these aspects of Jain's NCMT \cite{jain1} to frame a 
model which provides a better account for the {\it typical nature} of 
non-trivial dependence of $B$ on $N$ revealed from experimental 
observations on selected M:He$_N$ clusters. In this context it may be 
mentioned that our intial efforts \cite{sam} tried to use some 
simple thoughts to explain the effect by presuming that: (i) each 
added atom which takes the cluster from M:He$_N$ to M:He$_{N+1}$ 
can significantly change the positions of other $^4$He atoms from 
the axis of rotation, and (ii) with $N$ increasing beyond its certain 
value (depending on several physico-chemical aspects of M), $^4$He 
atoms start occupying the second position from M ({\it e.g.} in 
M(zero)-$^4$He(first)-$^4$He(second)) 
and these atoms interact so weekly with the rotor-part of cluster 
(M and few $^4$He atoms, -at first position, which interact directly 
with M) that they do not follow the rotation of the rotor.  In a sense 
the net potential seen by the rotor part of the cluster appears to 
remain constant with a change in the angular position of the rotor.  
In other words the rotor (when rotating about its axis) seems to role 
over a equi-potential surface; else if there are hills and valleys in 
the surface, the height of hills is much lower than the energy of 
rotational excitation of the rotor.  Although these efforts rendered a 
satisfying account of the phenomenon, they could not find a clear 
relation to superfluidity of $^4$He atoms and this motivated us in 
concluding this model which not only associates an important aspect of 
superfluid He-II with the non-trivial dependence of $I$ on $N$ but also 
clarifies how only few $^4$He atoms attached directly to M take part in 
the rotation and rests do not follow the rotation effectively. 

\bigskip
In what follows from the experimentally observed $N$ dependence 
of $B$ (represented typically by a curve depicted in Fig.(1)), 
we observe the following: 

\bigskip
\noindent
(A) $B$ decreases when $N$ is increased from $N=N^i$ (the lowest 
$N$ for which experimental data are available) to certain 
$N = N^*$ that may depend on several factors like the size 
and symmetry of the structure of M, the strength of M-He 
interaction, {\it etc}.
    
\bigskip
\noindent
(B) It remains nearly constant when $N$ is increased beyond 
$N^*$ only by 1 or 2 or so but increases with further increase 
in $N$ up to another value, say $N_1$.  

\bigskip
\noindent
(C) When $N$ is increased beyond $N_1$, $B$ is observed to 
decrease and increase alternately over different ranges of 
$N$, $N_1-N_2$, $N_2-N_3$, $N_3-N_4$, and $N_4-N_5$ and so on.

\bigskip
\noindent
(D) $\Delta\nu$ is observed first to increase linearly for 
first few $^4$He atoms (with $N < N^*$) but beyond this 
point it decreases with nearly a linear dependence on $N$; 
however, the slope of this decrease has different values 
over the ranges, 
$N^*-N_1$, $N_1-N_2$, $N_2-N_3$, $N_3-N_4$, and so on. 

\bigskip
It is well known that $\Delta\nu$ is a simple consequence of 
a change in potential $V(Q_1, Q_2, ... ... Q_S)$ (governing all 
the $S$ possible vibrational modes of M) with changing $N$.  
It could be explained in terms of a small change in the related 
potential constant appearing in the harmonic component in 
the expansion of $V(Q_1, Q_2, ... ... Q_S)$.  Although, it is 
difficult to argue in favour of increase or decrease in the value 
of potential constant of the chosen vibration, yet, however, a 
simple logic indicates that $\nu$ should increase for first few 
$^4$He atoms which occupy position in the closest 
vicinity of M since He-atoms are saturated with the electron 
charge density for which they would give away a small fraction 
of their own electron density to M which should strengthen the 
forces that govern its different modes of vibration and this 
is corroborated by experimental observation.  Although, with 
$^4$He atoms occupying second or third, ... positions counting 
from M (at zero-th position), it is difficult to argue whether 
$\nu$ would increase 
or decrease with $N$, however, it is clear the effect on $\nu$ 
should decrease with each added atom and this expectation agrees 
with decrease in slope of $\Delta\nu$ {\it vs} $N$ observed 
experimentally.  In what follows from these points, the change 
in $\Delta\nu$ with $N$ has nothing to do with microscopic 
superfluidity of the $^4$He atoms in the cluster. Hence, in 
this paper, we simply concentrate to find the origin of (A), 
(B) and (C) and conclude a general model of the phenomenon. 

\bigskip
\noindent
{\bf 1}. To a good approximation, the experimental observation 
of decrease in $B$ (or increase in $I$) for $N^i$ to $N^*$ 
can be explained by using rigid rotor picture for the cluster 
since this falls in line with the fact that a $^4$He atom 
interacts more strongly with M than with another $^4$He atom, 
the structure of the cluster M:He$_N$ for $N=N^i$ to $N=N^*$ 
can, therefore, be presumed to have a rigid rotor structure 
for the first few rotational excitations of each cluster. We 
note that $N^*$ can be different for different M (depending on 
its physico-chemical nature), while $N^i$ can, in principle, be 
as small as 1.  

\bigskip
\noindent
{\bf 2}. However, the non-trivial dependence of $B$ on $N$ 
represented by the observations that B remains nearly 
unchanged when $N$ changes by one or two $^4$He-atoms beyond 
$N^*$ and there after it follows cycles of its increase 
from $N^*+\Delta N$ to $N_1$, $N_2$ to $N_3$, ... and decrease 
from $N_1$ to $N_2$, $N_3$ to $N_4$, ... .  This unexpected 
observation, naturally indicates its relation with 
superfluidity of $^4$He-atoms.  Consequently, we try to 
explain it interms of an important aspect of superfluid 
$^4$He exhibited by it ubder the influence of its rotation. 
  
\bigskip
The fact, that different atoms on different concentric 
circles around the axis of a quantum vortex in He-II move 
coherently in order of their locations in a manner that they 
have no difference in their angular momentum \cite{wilks} 
implies that their angular velocity ($\omega$) changes as 
$r^{-2}$ where $r$ is the distance of the atom from the axis 
of the vortex.  This differs from atoms of a rigid body rotor 
where all atoms move around the axis of rotation with 
identically equal $\omega$.  In the following we consider the 
example of a set of $^4$He atoms moving on two concentric 
circles around the axis of rotation (as shown in Fig.2A) under 
the condition of: (i) constant angular velocity and (ii) constant 
angular momentum.  To this effect we 
evaluate the kinetic energy of the set by using, 
$$E = \sum_i^6\frac{n_i}{2}mr_1^2\omega_1^2 +  
\sum_j^{12}\frac{n_j}{2}mr_2^2\omega_2^2,  \eqno(1)$$

\noindent
where we have $n_i = 1, 2, 3, ...6$ and $n_j = 1, 2, 3, ...12$ 
with indices $i$ and $j$ to identify different atoms on orbits 1 
and 2, respectively.  Presuming that the radii of orbits 1 and 2 
satisfy  
$$r_2 = 2r_1 = 2r_o \eqno(2)$$

\noindent
and all atoms move as a single rigid body with  
$$\omega_1 = \omega_2 = \omega_o, \eqno(3)$$ 

\noindent
we have 

$$E = \frac{1}{2}\left[6mr_o^2\omega_o^2 + 
48mr_o^2\omega_o^2\right] = \frac{1}{2}\left[54mr_o^2\omega_o^2\right]  
= \frac{1}{2}I\omega_o^2 \quad \quad {\rm or} \quad I = 54mr_o^2. \eqno(4)$$ 

\noindent
Since the condition of constant angular momentum, applied to two atoms 
moving on different orbits of radius $r_1$ and $r_2$ 
({\it cf.}, Fig.2B) renders
$$mr_1^2\omega_1 = mr_2^2\omega_2  = C \quad ({\rm constant}) 
\quad \quad {\rm or} \quad 
\omega \propto r^{-2}, \eqno(5)$$

\noindent 
which implies that corresponding linear velocity $v = r\omega$ changes as 
$r^{-1}$.  As expected, this agrees with the well known dependence 
of $v$ on $r$ in a quantum vortex observed in superfluid $^4$He 
\cite{wilks}.  Using Eqn.(5) in Eqn.(1), we get 

$$E = \frac{1}{2}\left[6mr_1^2\omega_1^2 +  
12mr_2^2\omega_1^2\left(\frac{r_1}{r_2}\right)^4\right]  \eqno(6)$$

\noindent
which for the orbits satisfying Eqn.(2) and $\omega_1 = \omega_o$ 
renders
 
$$E = \frac{1}{2}\left[6mr_o^2\omega_o^2 + 
12mr_o^2\omega_o^2\frac{1}{4}\right] \quad =
\frac{1}{2}\left[9mr_o^2\omega_o^2\right]; \quad \quad 
{\rm with} \quad I = 9mr_o^2. \eqno(7)$$
      
\noindent
Eqn.(6) clearly reveals that the 
contribution to $I$ from an atom at a distance $r_2$, under the 
condition of constant angular momentum (Eqn. 5), gets reduced
 by a significant factor of $(r_1/r_2)^4$ (since $r_1 < r_2$) in 
comparison to that found under the condition of constant angular 
velocity (Eqn.(4)).  As an  example, the contribution to $I$ by an 
atom added to an orbit satisfying $r_2 = 2r_1$, Eqn.(2) gets 
reduced to a value as low as 1/16 ({\it i.e.} $\approx$6.3$\%$) 
and for the orbit satisfying $r_2 = 3r_1$, the said contribution 
is as low as 1/81 (or $\approx$1.3$\%$).  This speaks 
of the smallness of the contribution of an added atom to the 
$I$ of the cluster when it goes to an orbit of higher $r$ and 
evinces that the said atom has a $+ve$ contribution indicating 
that $I_{N+1} > I_N$.  

\bigskip 
In what follows the above stated inferences, we can expect a small 
increase or almost no change in $I$ for each added atom to the 
cluster with $N=N^*$.  However, it gives no clue for the experimental 
observations of $I_{N+1} < I_N$ for $N > N^*$.  To this effect our 
critical thinking reveals that the phenomenon is possible only if the 
added atom reduces the distance of $^4$He atoms (all the $N$ atoms or 
a few of them) in M:$^4$He$_N$ cluster from the axis of rotation to 
an extent that contribution to $I$ by the added single atom is over 
compensated by the decrease in $I_N$.   It is also possible if the 
added atom transforms the structure of M:$^4$He$_N$ cluster in a manner 
that one atom from first orbit of radius $r_1$ moves to the orbit of 
radius $r_2$ of the added atom (as shown in Fig.2C);  in this case 
the net change $\Delta I = I_{N+1} - I_N$ becomes 
$$\Delta I = 2mr_2^2\frac{r_1^4}{r_2^4} - mr_1^2 =
mr_1^2\left[2\frac{r_1^2}{r_2^2} -1\right] \eqno(8)$$ 

\noindent
which assumes a $-ve$ value for $r_2 > \sqrt{2}r_1$ indicating that 
$I_{N+1} < I_N$ when $r_2 > \sqrt{2}r_1$.  
Using this possibility for $N$ increasing beyond $N^*$ by 1 
atom, we have 

$$\Delta I = -0.5mr_o^2\eqno(9)$$ 

\noindent
by using Eqns.(2) and (8).  Presuming further that another atom moves 
similarly from orbit-1 to orbit-2 when an added atom to the 
cluster occupies orbit-2 as shown in Fig.2D, we have   
 
$$\Delta I = 4mr_2^2\frac{r_1^4}{r_2^4} - 2mr_1^2 =
mr_1^2\left[4\frac{r_1^2}{r_2^2} -2\right] = - 1.0mr_o^2 \eqno(10)$$ 

\noindent
for orbits satisfying Eqn.(2). Generalizing 
Eqns.(8) and Eqn.(10), we have  
 
$$\Delta I = mr_1^2\left[2n_c\frac{r_1^2}{r_2^2} - n_c\right]
 =  mr_o^2\left[\frac{2n_c}{\alpha^2} -n_c\right] \eqno(11)$$ 

\noindent
which represents the change in $I$ when $n_c$ atoms 
(above $N^*$) added to orbit-2 (making total $N = N^* + n_c$) 
induce $n_c$ atoms from orbit-1 to jump to orbit-2.  
Eqn.(11) reveals that $I$ has no change if 
$\alpha (=r_2/r_1) = \sqrt{2}$, it decreases by 
$\Delta I = -0.5n_cmr_o^2$ for $r_2 = 2r_1 = 2r_o$ 
({\it i.e.}, $\alpha = 2$) and by a maximum 
of $\Delta I = -n_cmr_o^2$ for $r_2 > > r_1 (= r_o)$. Such changes 
in $I$ for $n_c = 1, 2, 3, ...$ for different $\alpha = r_2/r_1$ 
are depicted in Fig.3 for their better perception.   

\bigskip
For a possible situation where no atom jumps from orbit-1 to orbit-2 
when $(n_c+1)$-th atom is added to orbit-2, it is evident that the 
added atom increases $I$ by $mr_o^2/\alpha^2$.  We have 
$$\Delta I =  mr_o^2\left[\frac{2n_c}{\alpha^2} -n_c\right] 
+ mr_o^2\frac{1}{\alpha^2}.  \eqno(12)$$

\noindent
which again means $I_{N+1} > I_N$ provided the added atom makes no
change in $\alpha$.  However, if it does and changes in $r_2$ and 
$r_1$ are such that $\alpha$ increases to $\alpha^* = \alpha + 
\Delta\alpha$, then by using Eqn.(12), we find 
$$\delta{(\Delta{I})} = \Delta I(\alpha^*) - \Delta I(\alpha) 
= -mr_o^2\frac{2(2n_c+1)\Delta\alpha}{\alpha^3}. \eqno(13)$$ 

\noindent
We note that this $-ve$ change in $I$ can overcompensate the 
increase in $I$ by $mr_o^2/\alpha^2$ if 

$$\Delta\alpha \ge \frac{\alpha}{2(2n_c+1)} \quad \left(= 
\frac{1}{2(2n_c+1)}\frac{r_2}{r_1}\right)\eqno(14)$$  

\noindent
which is obtained by equating RHS of Eqn.(13) to $ mr_o^2/{\alpha^2}$.  
This indicates that $I$ of the cluster can have continuous 
decrease with increase in $N$ (possibly from $N^*$ to $N_1$) 
if $\alpha = r_2/r_1$ increases by an appropriate value of 
$\Delta\alpha$ with each added atom. In other words an agreement 
between theory and experiment can be seen by using $n_c$, $\alpha$, 
and $\Delta\alpha$ as adjustable parameters.  Note that increase 
in $\alpha$ is possible both by decrease in $r_1$ and increase 
in $r_2$ when an atom is added to the cluster. However, it 
appears that desired increase in $r_2$ is more probable than 
decrease in $r_1$.   As revealed by Eqn.(14), decrease in $I$ 
is possible if $\Delta\alpha$/$\alpha$ increases by more than 
16.6$\%$, 10$\%$ and 7.1$\%$, respectively, in case of 
$n_c =$ 1. 2, and 3. These aspects are depicted in Fig.4 for their 
better understanding. 

\bigskip
Summing up the possible explanation for the phenomenon in the 
light of our preceding analysis, we may mention that : 

\smallskip
\noindent
(1) With increasing $N$ from $N^i$ to $N^*$, $^4$He atoms in M:$^4$He$_N$ 
cluster (for $N \le N^*$) seem to have reasonably strong binding 
with M for which the cluster as whole represents a rigid rotor, 
to a good approximation, and its $I$ increases (or corresponding 
$B$ decreases) with $N$ in agreement with experiments ({\it cf.} Fig.1, 
for $N \le N^*$). 

\smallskip
\noindent
(2) For nearly no change in $I_{N}$ from $I_{N^*}$, when $N$ is 
set to have a value $N^* + \Delta N$ (where $\Delta N$ has only 
small value such as 1 or 2 or so), it appears that each of the 
$\Delta N$ atoms go to orbit 2 for which 
$\Delta I (= I_{N^*+1}- I_{N^*}$) is $\approx$ 6.3$\%$ of the 
contribution of an atom in orbit 1.  Such a small increase can be 
easily compensated if the contribution to $I_{N^*+1}$ due to 
each of the $N^*$ atoms in orbit 1 gets reduced by 6.3/$N^*\%$ 
presumably due to small decrease, $\Delta r$, in the distance of 
atoms from the axis of rotation of M:$^4$He$_{N^*}$ and this does 
not demand necessarily a decrease in M-$^4$He bond length; a 
decrease by $\Delta r$ in the projection of the bond on the plane 
$\perp$ to the axis of rotation would suffice and this can be 
estimated to fall around 6.3/2$N^*\%$ of the said projection 
which equals to $\approx 0.8\%$ if $N^*= 4$, or 0.5$\%$ if $N^* = 6$ or 
$\approx 0.4\%$ if $N^* = 8$.  Such a small change can easily 
be expected as a possible effect of an atom added to M:$^4$He$_{N^*}$ 
or another added to M:$^4$He$_{N^*+1}$; this naturally explains the 
said observation. 

\smallskip
\noindent
(3) For the remaining part of $I_N$ {\it vs.} $N$ curve ({\it i.e.}, 
for $N > N^*+\Delta N$) where $I_N$ is observed to have significant 
decrease with increase in $N$, changes in $\alpha = r_2/r_1$ along 
with the jump of an atom from orbit 1 to orbit 2 seem to take place 
when an atom is added to orbit 2. Depending on the physico-chemical 
nature and size of M, the decrease in $I_N$ for $N$ changing from 
$N=N^*+\Delta N$ to $N=N_1$ can be explained by choosing $n_c =1$ 
and 2 ... (in different steps) clubbed with appropriate values of 
$\alpha$ and $\Delta\alpha$.  This speaks of the sensitivity of 
the changes in $I_N$ on $\alpha$ and $\Delta\alpha$ as well as 
$\Delta r$ (change in the said projection of M-$^4$He bond length).  
This naturally simplifies the basis our understanding of the 
observed increase and decrease in $I$ (Fig.1) with $N$. (4) The 
decrease in $B$ (increase in $I$) for $N$ increasing from $N_1$ 
to $N_2$ is as per normal expectation.  However, each atom added 
to the cluster in this range contributes only very small fraction 
of the contribution to $I$ by an atom in orbit 1. Guided by this 
fact it appears that value of $\alpha$, $\Delta\alpha$ and 
$\Delta r$ should explain the $N-$ dependence of $I_N$ not only 
in this range but for all values of $N > N_1$.    

\bigskip
Identifying the ranges, $0- N_1$, $N_1-N_2$, $N_2-N_3$, 
{\it etc.} (Fig. 1) as cycles of inccrease and decrease in $I_N$, it 
appears that M:$^4$He$_N$ cluster has different shells of $^4He$ 
atoms around M and each cycle represents the completion of one 
shell.  While $^4$He atoms in the first shell have direct bond with M, 
those in second, third, ... shells are separated from M, respectively, 
by 1, 2, ... $^4$He atoms in between.  The maximum number of $^4He$ 
atoms, in a particular shell increases in proportion of $R^2$ (where $R$ 
is the radius of the shell which changes from one shell to next shell 
in units of the diameter of the sphere which represents the shape and 
size of $^4$He atom; however it also depends on the shape and size 
of M.  Assuming that M has a shape and size of a $^4$He atom, a rough 
estimate reveals that the number of $^4$He atoms in first, second 
and third shells, should be {\it around} 6, 18 and 40, respectively.  
However, it may be emphasized that these numbers agree approximately 
with experimental values because M may have linear or a complex 
structure.  Further, it may also be mentioned that it is not the 
length of M-$^4$He bond which changes much with atoms added to the 
cluster but the projection of this bond on the plane $\perp$ to the 
axis of the rotor which should be considered to explain the changes 
in $I_N$ with added $^4$He atoms that we observe through experiments.

\bigskip
The experimental observations seem to indicate that the part of 
M:$^4$He$_N$ cluster which rotates, to a good approximation as a 
rigid rotor, has fewer than $N^*$ $^4$He atoms in clusters of 
$N >> N^*$.  This agrees with our suggestion that $n_c$ 
(= 1 or 2, or so) atoms move from orbit 1 to orbit 2 with increasing 
$N$ beyond $N^*$.  However, the physics of this possibility is not 
yet very clear.  

\bigskip
In the light of the fact that $^4$He atoms in superfluid state make a
close packed arrangement of their wave packets (CPA-WP) supported by a 
number of experimental observations such as the observation of Stark 
effect of roton transition seen through microwave absorption 
\cite{jains} and the unequivocal conclusions of a number of theoretical 
studies \cite{jain2, jain1, simanta}, the rotor part of the 
cluster in CPA-WP type arrangement of $^4$He atoms may in certain cases 
experience a kind of low energy potential 
barrier with an axial symmetry of the order $n$ (as shown in Fig.5).  
A theoretical analysis for such a case \cite{jain3} reveals that the 
effective $I$ of the rotor has a lower value that depends on the 
height and symmetry 
of $V_n$ ({\it cf.}, Fig. 5).  This renders an additional reason for 
a small decrease in $I$.     

\bigskip
As concluded by Jain's NCMT of a system of interacting bosons 
such as liquid $^4$He \cite{jain1}, atoms in He-II not only move coherently 
in order of their locations but also have identically equal 
angular momentum when they move on different concentric paths 
around the axis of a quantum vortex; in fact this theory for the 
first time answers a question raised by Wilks \cite{wilks} in relation 
to Feynman's account for the origin of quantized circulation 
\cite{feyn1}.  Wilks has rightly argued that Feynman's basis for quantum 
vortices in He-II is equally 
valid for He-I but the latter does not show any quantum vortex.  Using 
a basic aspect of quantum vortices observed in He-II, we discover a 
model which has enough potential to explain qualitatively the typical 
nature of experimentally 
observed $N$ dependence of the rotational constant $B$ of the rotor 
part of the cluster M:He$_N$.  Naturally, the question, how exactly 
superfluidity is related to the said dependence of $B$ is answered 
with utmost clarity.  We hope that this model, when used with simulation 
techniques on individual cluster, would render results that would 
agree closely with experiments.  This would not only help in improving 
the model but also for having a clear understanding of the phenomenon.  
However, we could not take up this task for the want of facilities of 
computer simulations at our end.        

{}

\begin{figure}[H]
\begin{center}
\includegraphics[angle = 0, width=1.0\textwidth]{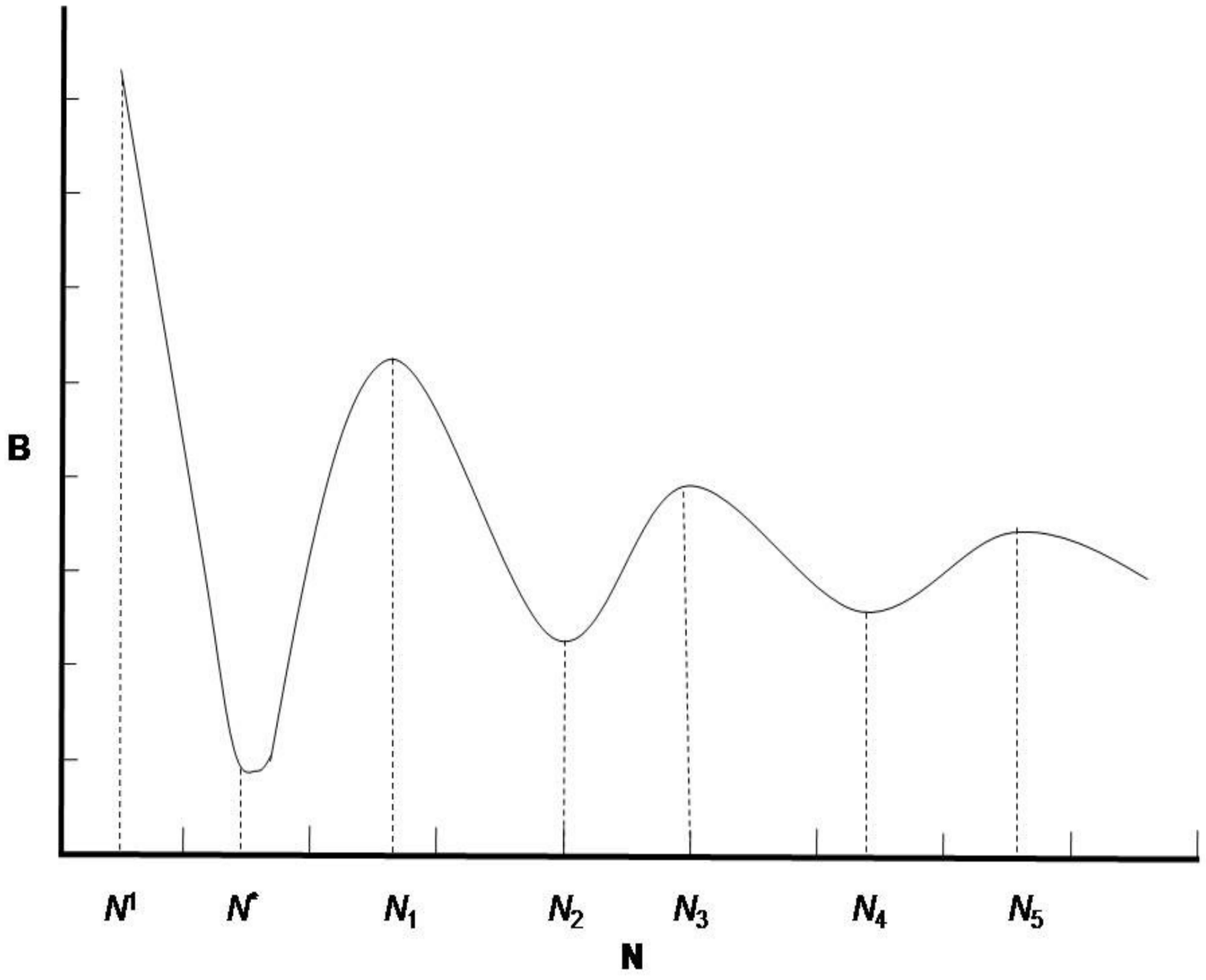}
\caption{Typical nature of the $N$ dependance of rotational constant 
(B) of the rotor in a M:He$_N$ cluster. }
\end{center}
\end{figure} 
          
\begin{figure}[H]
\begin{center}
\includegraphics[angle = 0, width=1.0\textwidth]{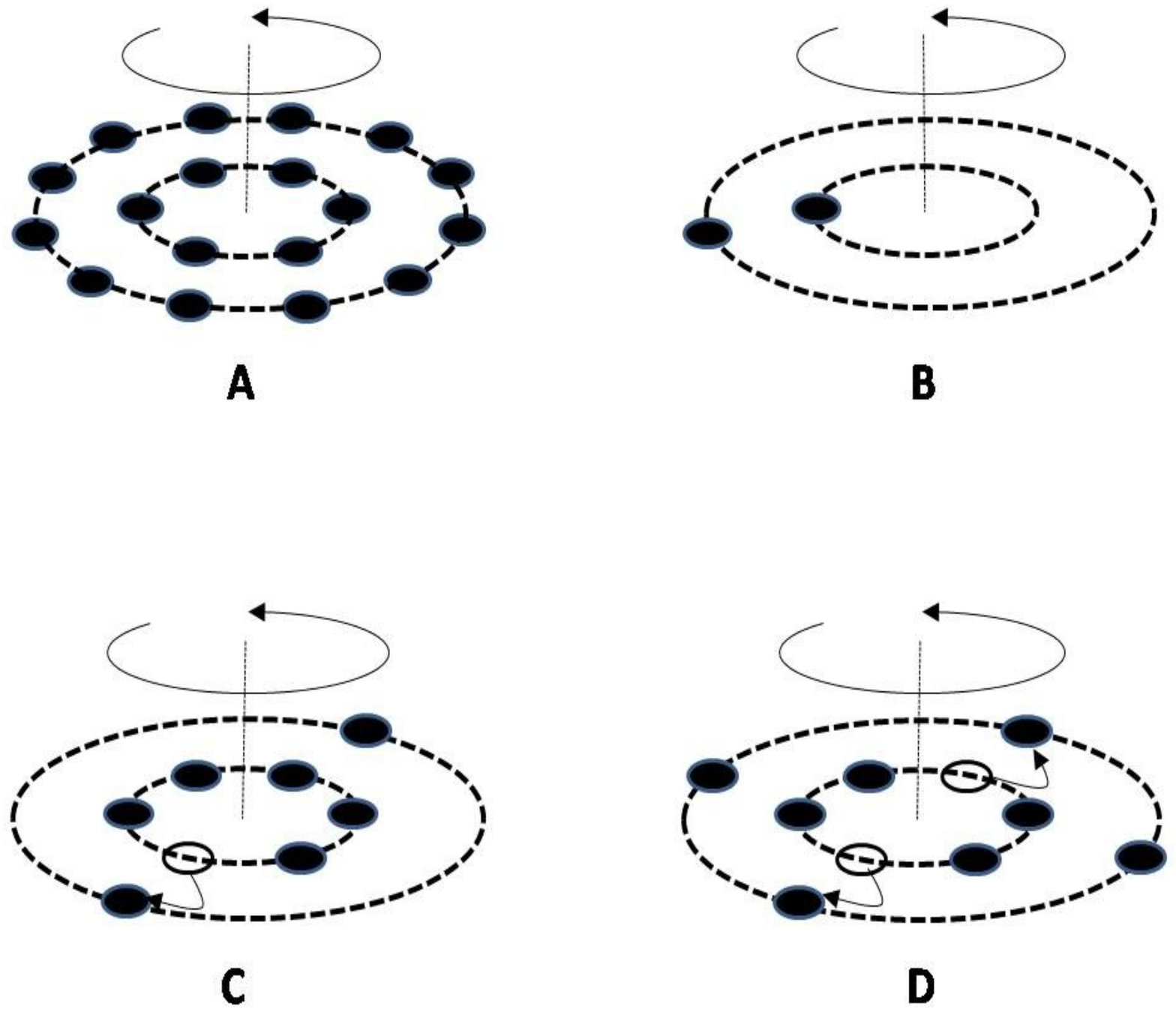}
\caption{(A) Two concentric circular paths of a vortex on which atoms 
move in order of their locations, (B) two atoms moving on different 
concentric paths, (C) jump of an atom from inner path to outer path 
with an added atom in the cluster (D) jump of two atoms from inner 
path to outer path with two added atoms in the cluster}
\end{center}
\end{figure} 

\begin{figure}[H]
\begin{center}
\includegraphics[angle = 0, width=1.0\textwidth]{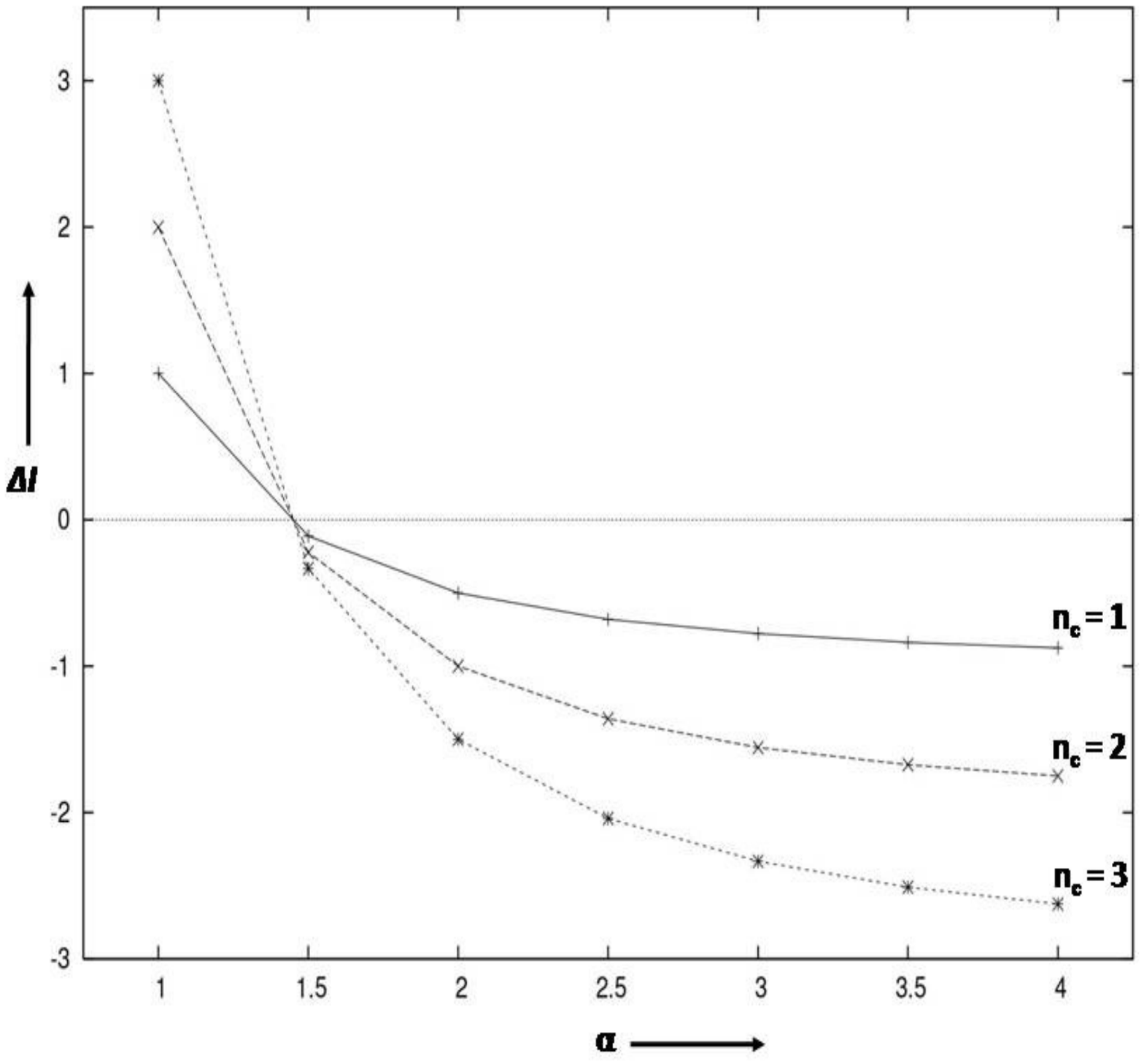}
\caption{Dependence of $\Delta I$ on $\alpha = r_2/r_1$ (Eqn.11) 
for $n_c =$ 1, 2, and 3}
\end{center}
\end{figure}

\begin{figure}[H]
\begin{center}
\includegraphics[angle = 0, width=1.0\textwidth]{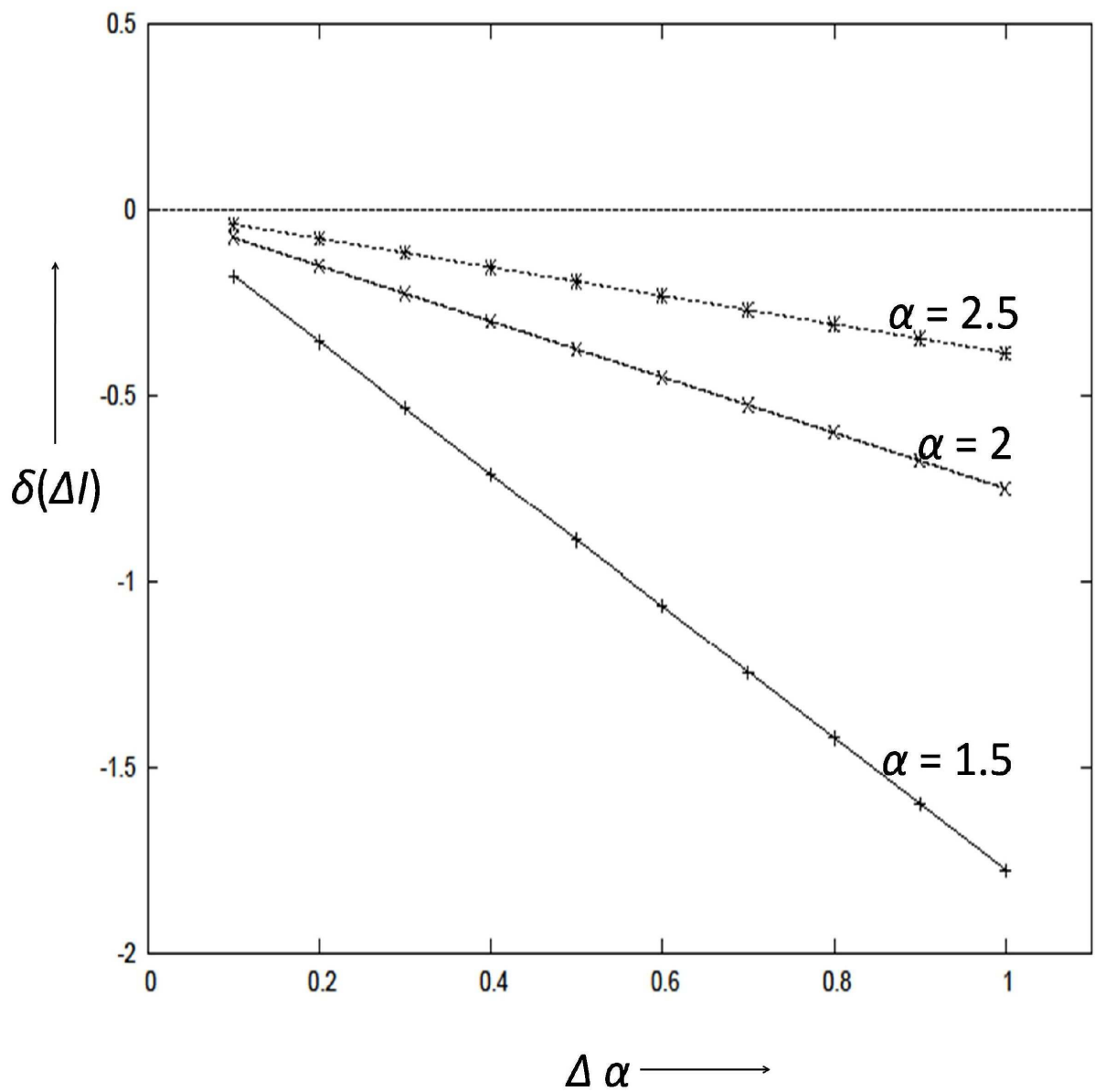}
\caption{Dependence of $\delta{(\Delta I)}$ on $\Delta\alpha$ (Eqn.13) 
for $n_c = 1$ and different values of $\alpha$}
\end{center}
\end{figure} 

\begin{figure}[H]
\begin{center}
\includegraphics[angle = 0, width=1.0\textwidth]{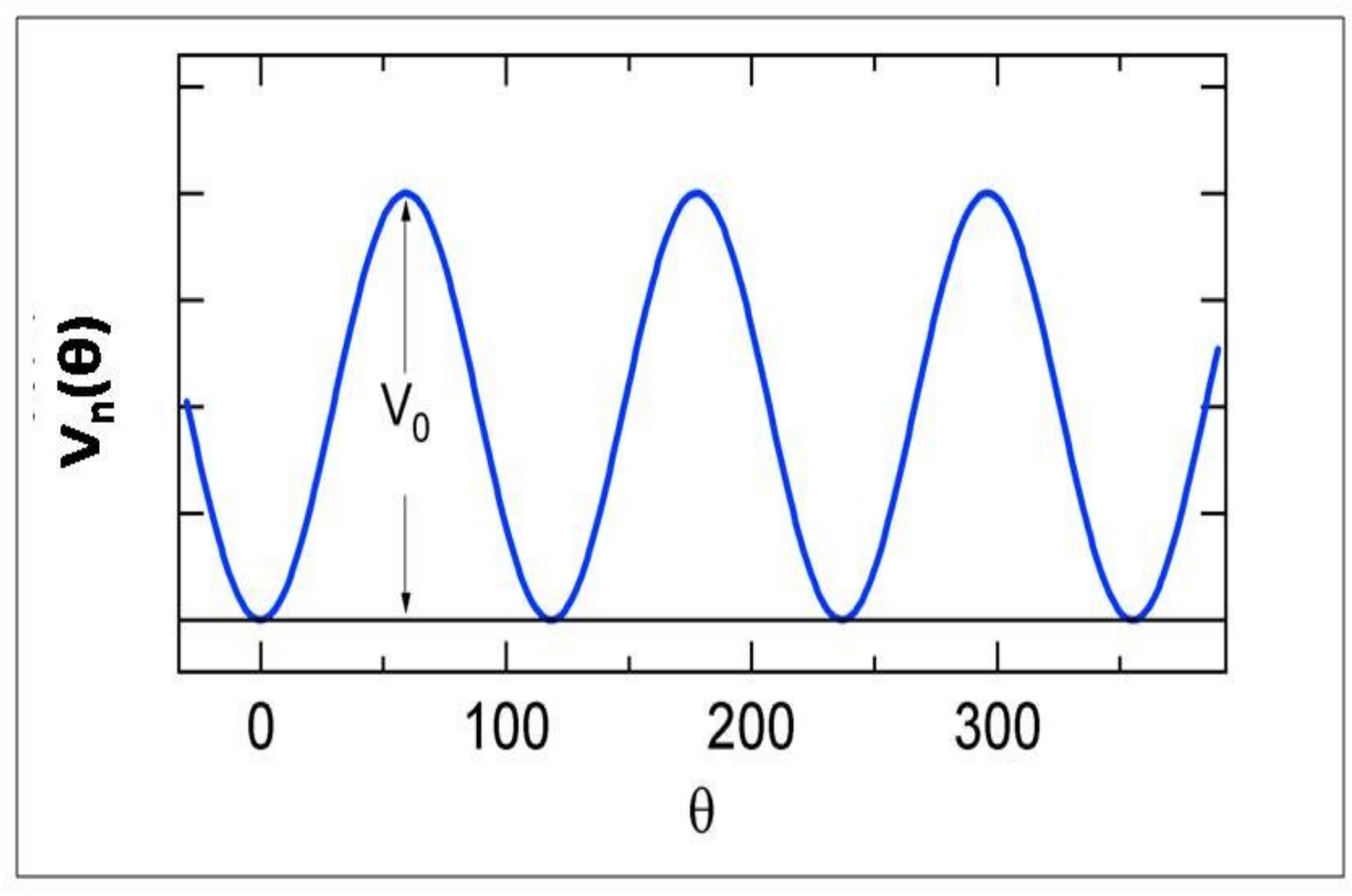}
\caption{Variation of n fold symmetric potential with rotation angle 
$\theta$}
\end{center}
\end{figure} 

\end{document}